\documentclass[useAMS, twocolumn, aps, prd, amscd, amsmath, amssymb,verbatim,nofootinbib,usenatbib]{revtex4-1}
\usepackage{times}
\usepackage{graphicx}
\usepackage{verbatim}
\usepackage{tikz}
\usepackage{color}

\usepackage[T1]{fontenc}
\usepackage{aecompl}
\usepackage{calligra}
\DeclareMathAlphabet{\mathcalligra}{T1}{calligra}{m}{n}
\DeclareFontShape{T1}{calligra}{m}{n}{<->s*[2.2]callig15}{}

\usepackage{hyperref}
\hypersetup{
    pdfnewwindow=true,      % links in new window
    colorlinks=true,       % false: boxed links; true: colored links
    linkcolor=blue,          % color of internal links
    citecolor=cyan,        % color of links to bibliography
    filecolor=blue,      % color of file links
    urlcolor=blue           % color of external links
}

\def\prd{PRD}
\def\apj{ApJ}                 % Astrophysical Journal
\def\apjl{ApJL}               % Astrophysical Journal, Letters
\def\mnras{MNRAS}             % Monthly Notices of the RAS
\def\aap{A\&A}                % Astronomy and Astrophysics
\def\nat{Nature}              % Nature
\def\aapr{A\&A Review}

\def\bX{{\bf X}}
\def\bS{{\bf S}}
\def\hn{\hat {\bf n}}
\def\hnc{\hat n}
\def\hk{\hat {\bf k}}
\def\hs{\hat {\bf s}}
\def\tA{\tilde{A}}

\def\obs{\mathrm{obs}}

\def\min{\mathrm{min}}
\def\max{\mathrm{max}}

\def\Msun{{M_{\odot}}}

\def\GW{\mathcal{GW}}
\def\Mc{{\mathcal{M}}}
\def\H{{\mathcal{H}}}

\def\be{\begin{equation}}
\def\ee{\end{equation}}

\def\bea{\begin{eqnarray}}
\def\eea{\end{eqnarray}}

\def\eg{\textit{e.g.}}

\begin{document}

\title[Gravitational Wave Parallax]{Using Gravitational Wave Parallax to Measure the Hubble Parameter with Pulsar Timing Arrays}

\author{Daniel J. D'Orazio}
\email{djdorazio@gmail.com}
%\email{daniel.dorazio@cfa.harvard.edu} %not valid after October 14, 2020
\affiliation{Department of Astronomy, Harvard University, 60 Garden Street Cambridge, MA 01238, USA}
\affiliation{Niels Bohr International Academy, Niels Bohr Institute, University of Copenhagen, Blegdamsvej 17DK-2100 Copenhagen, Denmark}

\author{Abraham Loeb}
\affiliation{Department of Astronomy, Harvard University, 60 Garden Street Cambridge, MA 01238, USA}

\begin{abstract}
We demonstrate how pulsar timing arrays (PTAs) can, in principle, yield a purely gravitational
wave (GW) measurement of the luminosity distance and comoving distance to a
supermassive black hole binary source, hence providing an estimate of the
source redshift and the Hubble constant. The luminosity distance is derived
through standard measurement of the chirp mass, which for the slowly evolving
binary sources in the PTA band can be found by comparing the frequency of
GW-timing residuals at the Earth compared to those at distant pulsars in the
array. The comoving distance can be measured from GW-timing parallax caused
by the curvature of the GW wavefronts. This can be detected for single sources 
at the high-frequency end of the PTA band out to Gpc distances with a future PTA
containing well-timed pulsars out to $\mathcal{O}(10)$~kpc, when the
pulsar distance is constrained to less than a GW wavelength. 
Such a future PTA, with $\gtrsim 30$ pulsars with precise distance measurements
between 1 and 20~kpc, could measure the Hubble constant at the tens of percent level
for a single source at $0.1 \lesssim z \lesssim 1.5$.  At $z\lesssim0.1$, the
luminosity and comoving distances are too similar to disentangle, unless the
fractional error in the luminosity distance measurement is decreased below
$10\%$. At $z\gtrsim 1.5$, this measurement will likely be restricted by a
signal-to-noise ratio threshold. Generally, clarification of the different
types of cosmological distances that can be probed by PTAs, and their relation
to pulsar distance measurements is important for ongoing PTA experiments aimed
at detecting and characterizing GWs.
\end{abstract}

\maketitle

\section{Introduction}
Gravitational waves (GWs) from coalescing compact object binaries
are now being used to measure cosmological parameters. This new
handle on cosmology is an important tool for understanding systematics in
our current measurements of the cosmological parameters, \eg, for
resolving the existing tension between different measures of the Hubble
constant \citep{Riess+2016, Riess+2019, PlanckH0:2016}.

Direct measurements of the Hubble constant rely on knowledge of the redshift
of an emitting source in addition to a determination of its intrinsic
luminosity, be it GW or electromagnetic (EM), which is used to determine the
luminosity distance. Comparison of the redshift and distance yields the Hubble
constant. For example, the backbone of the standard candle approach
\citep{Riess+2019} leverages the Leavitt Law \citep{LeavittLaw:1912} to relate
the oscillation period of Cepheid variable light-curves to the intrinsic
luminosity, while a redshift is measured from the frequency shift of spectral
lines in the host galaxy. The standard sirens approach \citep{Schutz:1986,
KrolakSchutz:1987, HolzHughes:2005, CutlerHolz:2009, StandSirenGW170818:2017,
ChenHolzFishbach:2018} uses the predicted GW strain and frequency evolution of
a coalescing binary to determine a luminosity distance, while again relying on
an EM determination of the redshift, $z$.

While both are vital techniques, contributing independent measures of
cosmology, the former relies on theoretical knowledge of standard candles and
their astrophysical environments, \eg, supernovae,
\citep{Rigault:SNenvirH0:2018} while the latter relies on the existence of an
EM counterpart that can be identified with the GW source, and hence
understanding EM emission mechanisms. Additionally, both approaches only apply
out to distances where EM emission can be detected. 

The few methods that have been proposed to make cosmological measurements with
GWs alone make use of inferred knowledge of the rest-frame GW source
properties. For example, \citep{TaylorGair_DCI:2012,
TaylorGairMandel_DCII:2012} rely on models of the rest frame neutron star mass
distribution to break the mass-redshift degeneracy of in-spiraling neutron
star binaries. This is required because of the scale invariance of the binary
merger problem. The GW luminosity is independent of the binary mass (which is
why a luminosity distance can be measured), and the quantity $\Mc f$, the
chirp mass times the GW frequency, is invariant with redshift. Hence, one must
obtain knowledge about intrinsic source properties to make a joint redshift
and luminosity distance determination, or one must introduce a new scale to
the problem.

Here we do the latter. We propose a method for probing the distance-redshift relationship, and
hence, measure the Hubble constant, via gravity alone, and without making
assumptions about the GW source. While we cannot break the scale invariance of
the binary merger problem, we can infer the redshift through a different
means, with a large detector whose components are separated widely enough to
detect the GW wavefront curvature. Measurement of this curvature through
timing parallax will provide a distance to the source that is formally a
comoving distance. The comoving distance, $D_c$, is related to the luminosity distance, $D_L$, through $D_c = (1+z)^{-1} D_L$. A separate determination of
the luminosity distance from the GW chirp and amplitude gives the redshift.
Comparison of redshift and luminosity distance yields the Hubble constant up to choices of the cosmological density parameters.

Such a determination of the comoving distance from GW wavefront curvature is
not possible with current interferometric GW detectors such as LIGO
\citep{LIGO} and LISA \citep{LISA:2017}, which are sensitive to GWs from compact-object binary mergers ranging in mass
from a few to $\sim 10^7 \Msun$. However, Ref. \cite{DengFinn:2011} (hereafter DF11) shows
that it is possible with the galaxy-scale Pulsar Timing Arrays \citep[PTAs,][]{PTAs}, which
are expected to detect low frequency GWs from the biggest black hole mergers
in the universe with masses of $10^8-10^{10} \Msun$, within the next decade
\citep{Kelley_SS+2018}. We re-derive this result with the important
clarification that the distance recovered in this manner is indeed a comoving
distance, not a luminosity distance as posited in DF11.

\section{Distance Measurements with PTA{\scriptsize s} }
\label{S:Distance Measurements with PTAs} 
The PTAs employ millisecond pulsars (MSPs) across the galaxy as precise
clocks. A GW passing through the Earth-pulsar array will cause detectable
deviations in the arrival times of the otherwise steady pulses that, when meticulously separated from non-GW induced timing residuals due to intrinsic changes in pulsar period and the intervening Earth-pulsar medium \citep{PTAnoise:2018}, will allow detection of GWs in the 1-100~nHZ frequency band. This is the relevant band for tracking the late inspiral of the most massive, $10^8-10^{10} \Msun$ black holes binaries at the hearts of massive galaxies \citep{BurkeSpolar:nHzGW+2019}.

Unlike their high-frequency interferometric-detector cousins, LIGO and LISA,
one of the PTA's primary targets is a stochastic background of GWs, an
astrophysical noise floor generated by the superposition of many inspiraling
supermassive black hole binaries across cosmic distance \citep{SiemensGWB+2013C, TaylorGWBtime+2016}. Above this noise floor, it
is expected that a number of single resolved binary sources will also be
detected, where different population models place this number at $\mathcal{O}$(1-10) for near future arrays \citep{SVV:2009, Ravi:2014, Rosado:2015,
Kelley_SS+2018}. Here, we focus on the single resolved binary sources and show
how a luminosity distance and a comoving distance can be measured for a
subset of them.

\subsection{Luminosity Distance} 
\label{S:Luminosity Distance Measurement} 
It is well known that the luminosity distance can be measured for binary GW sources when their frequency evolution can be detected \citep[\eg,][]{Schutz:1986, StandSirenGW170818:2017}. So called chirping binaries allow measurement of the chirp mass $\Mc$ and the GW strain $h$. In the source frame these are related by
\begin{equation}
\label{Eq:hDc}
h \propto \frac{\Mc^{5/3}_s f^{2/3}_s}{D_c},
\end{equation}
where $s$ denotes the source frame and $D_c$ is the comoving distance.
Because the strain $h$ and GW frequency $f$ can be measured over time, and the chirp mass can be measured from the first time derivative of the frequency \citep[or the chirp, see][]{HolzHughes:2005,
DLGWlens:2020}, Eq. (\ref{Eq:hDc}) allows a measurement of the distance. In the observer's frame, the redshift of the frequency and its derivative implies that the strain, written in terms of observables, reads
\begin{equation}
h \propto \frac{\Mc^{5/3}_o f^{2/3}_o}{D_L},
\end{equation}
where $o$ denotes the observer's frame and the measurable distance is $D_L = (1+z)D_c$, the luminosity distance.

Whether or not a binary is chirping in the detectable band is set by the
timescale for GW frequency evolution. For a binary on a circular orbit, 
\begin{eqnarray}
\label{Eq:tchirp}
t_{\mathrm{chirp}} \sim \frac{f}{\dot{f}} &=& \frac{5}{96 } \left(\frac{G \Mc}{c^3} \right)^{-5/3} (\pi f)^{-8/3}  \\
&\approx& 5.43\times 10^3 \mathrm{yr}  \left(\frac{\Mc}{ 10^9 \Msun} \right)^{-5/3} \left(\frac{f}{ \mathrm{yr}^{-1}} \right)^{-8/3}. \nonumber
\end{eqnarray}
For the high frequency and low mass binary inspirals and mergers detected by
LIGO ($M=1-10^3\Msun$, $f=10-10^4$~Hz), and in the future, by LISA
($M=10^2-10^7\Msun$, $f=10^{-4}-10^{-1}$~Hz), $t_{\mathrm{chirp}}$ is short
compared to observation times and determination of the chirp mass and
luminosity distance is expected. For the PTAs, however, Eq. (\ref{Eq:tchirp})
shows that the time for the GW frequency to evolve in the PTA band can be
thousands of years. Hence, it is often assumed that only the combination
$\Mc^{5/3}_o/D_L$ is measurable for binary GW sources in the PTA band.

However, a number of works discussed below have pointed out that chirp
information in the PTA band can be gleaned by incorporating the many thousand
year light travel time across the Earth-pulsar detector. Because the timing
residuals measured on Earth are a culmination of the entire path traveled by a
EM pulse between the pulsar and Earth, the chirp can be detected by comparing
the GW signal at the pulsar (pulsar term) compared to the signal at Earth
(Earth term). Chirp detection then requires that the change in GW frequency at
the detector (Earth-pulsar system), over the course of the light travel time
across the detector, be larger than the frequency resolution of the detector.
Conservatively, the frequency resolution is given by the inverse of the
observation time, $\Delta f = 1/t_{\obs}$ \citep[\eg,][]{SesanaVecchio:2010,
Cornish:2003}. Hence the condition on the Earth-pulsar distance $L$ needed to
measure the GW chirp is,
\begin{equation}
L \geq \frac{c}{\dot{f} t_{\obs}} \approx 0.08 \mathrm{kpc} \ \left(\frac{\Mc}{ 10^9 \Msun} \right)^{-5/3} \left(\frac{f}{ \mathrm{yr}^{-1}} \right)^{-11/3} \left( \frac{t_{\obs}}{20 \mathrm{yr}} \right)^{-1}.
\label{Eq:DLcond}
\end{equation}

For standard pulsar distances in present day PTAs of $0.1-1$~kpc, this
condition is met for the high frequency, high mass end of the expected
binary population detectable by the PTAs. Longer pulsar baselines of future
arrays, reaching out to 20 kpc (see \S \ref{S:Distance Measurement Precision
and PTA Dependence}), could allow chirp detection from the largest, $10^{10}
\Msun$, binaries down to a few 10's of nHz, or from the smaller,
$10^{8}\Msun$,  binaries at $\sim100$ nHZ. 

The parameter space of binaries that meet this criterion and have a detectable
strain is explored further in Ref. \cite{Lee+ssPTA:2011}, while Ref.
\citep{Taylor:EccPTA+2016} discusses the region of binary parameter space
where assuming zero-frequency evolution could be detrimental for detection.
Ref. \cite{SesanaVecchio:2010} uses the synthetic population of supermassive black hole binaries from
Ref. \cite{SVV:2009} and shows that the majority will have resolvable chirps
when taking into account the pulsar term. While both studies point this out,
the primary focus of these works is not binary frequency evolution, and so the
chirp was ignored.  Ref.~\cite{CorbinCornish:2010}, however, investigates
recovery of the luminosity distance with PTAs from such pulsar-term, chirping
binaries. They assume an $\mathrm{SNR}=20$ detection, with 20 pulsars each having a
$100$~ns timing residual and randomly oriented on the sky at distances between
$0.5-1$~kpc. They find that the fractional error on the distance can be as low
as $7\%$ for edge-on inclination binaries ($i=\pi/2$) and rises to $30 \%$ for
$i=\pi/4$.  We take the findings from the above studies as conservative
estimates of how well the luminosity distance can be recovered as they each
assume pulsar distances on the order of 1~kpc, where, as motivated in the next
section, we are interested in more futuristic PTAs that contain well-timed
pulsars out to 20~kpc.

\subsection{Comoving Distance}
\label{S:Comoving Distance Measurement}
We now show for the first time how PTA observations of GWs from a resolved,
single-binary source can independently measure the source \textit{comoving
distance}.

\subsubsection{Geometrical Argument}
\label{S:Geometrical Argument}
The amount by which the arrival time of the EM pulses to Earth
deviates due to a passing GW is dependent upon the changing amplitude,
frequency and phase of the GW encountered by the EM pulse at it traverses the
Earth-pulsar distance. This is dependent on the shape of GW wavefronts,
surfaces of constant GW phase, across the Earth-pulsar system. For very
distant GW sources, the GW wavefronts can be assumed to be planar. However,
for nearby sources, the true spherical nature of the wavefronts becomes
non-negligible and encodes the source distance.

\begin{figure}
\begin{center}$
\begin{array}{c}
 \includegraphics[scale=0.25]{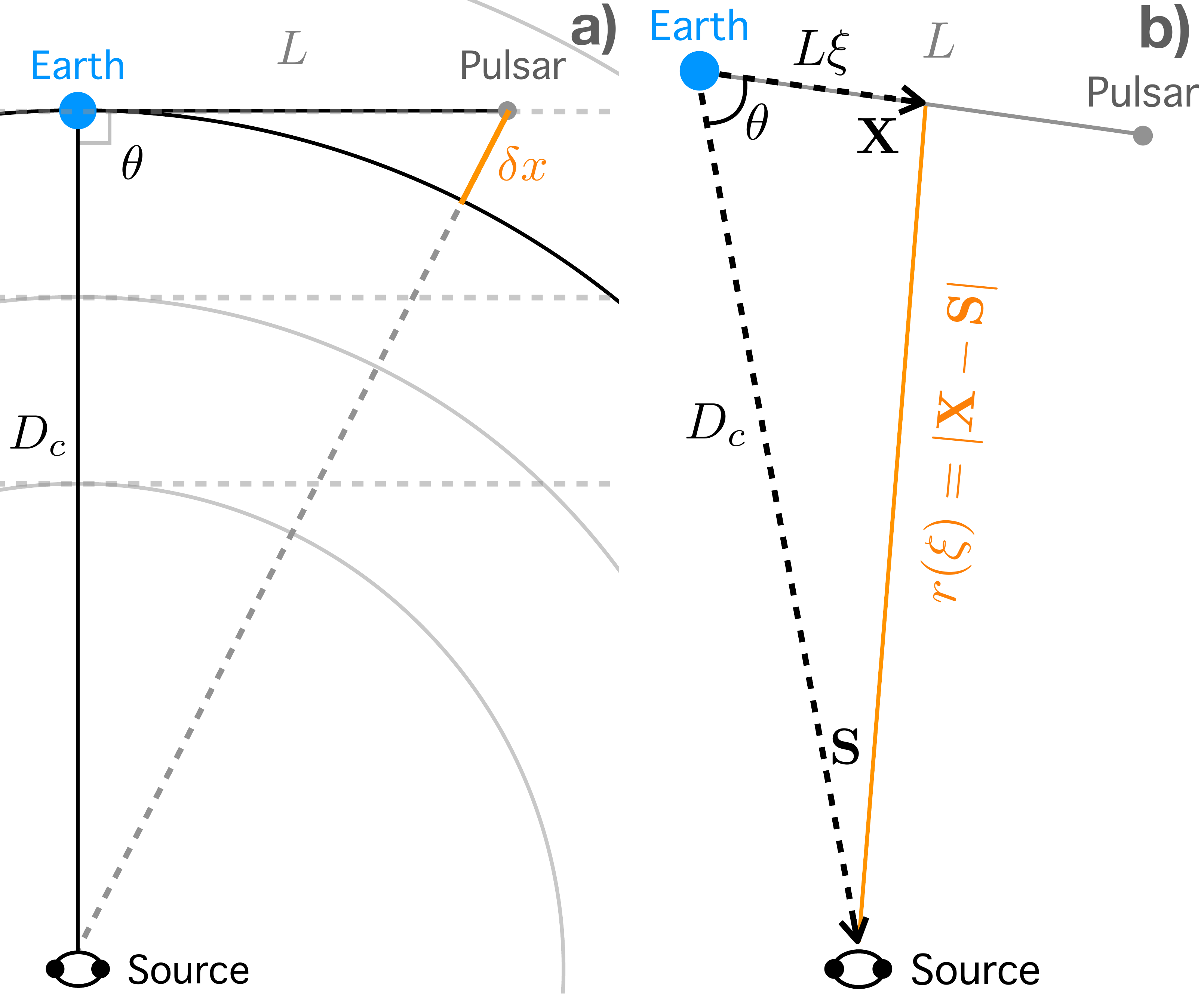}
\end{array}$
\end{center}
\vspace{-10pt}
\caption{
Schematics for visualizing the geometrical (panel a), \S \ref{S:Geometrical Argument}) and mathematical (panel b), \S \ref{S:Mathematical Argument}) description of the comoving distance measurement. In panel a), we assume $\theta=90^{\circ}$ to simplify the geometrical argument.
}
\label{Fig:schem}
\end{figure}

Panel a) of Figure \ref{Fig:schem} illustrates a geometrical argument that
elucidates this concept and provides an estimate of when wavefront curvature
is important. Our setup consists of the detector: Earth and a pulsar separated
by a distance L aligned at an angle $\theta$ relative to the line of sight of
a source of GWs with observed frequency $f$ at comoving distance $D_c$.

Without loss of generality, we consider the case where, under the plane-wave
approximation, GWs emitted at some time in the source frame arrive at the
Earth and the pulsar after the same travel time. For spherical wavefronts, the
travel time to the pulsar differs by $\delta t = \delta x/ c$ (panel a) in Figure \ref{Fig:schem}), causing an EM pulse to traverse a different accumulated GW phase along its path.
In comoving (flat-space) coordinates, we can compute $\delta t$ via Euclidean geometry,
\begin{equation}
\delta t = \frac{D_c}{c} \left[\sqrt{1 + \left(\frac{L}{D_c}\right)^2}  - 1\right] 
\approx \frac{1}{2} \left( \frac{L}{D_c} \right) \frac{L}{c}.
\end{equation}
This extra travel time compared to the plane-wave case only affects the pulse arrival time if the EM pulse encounters a significant extra portion of a GW cycle. Hence, a condition on there being a significant difference between timing residuals in the plane-wave and spherical-wave cases is found from requiring that $\delta t f \gtrsim 1$ (noting that at exact integer values there is no change). This places a limit on the distances $D_c$ and $L$ for which wavefront curvature is important,
\begin{equation}
D_c \lesssim f \frac{L^2}{c} = \left(\frac{L}{\lambda_{\GW}}\right) L,
\label{Eq:GeoCrit}
\end{equation}
where $\lambda_{\GW} = c/f$, in analogy to the Fresnel condition in optics.

Because our observable is influenced by the relative time of arrival of GW wavefronts across the Earth-pulsar system, the distance to the source, $D_c$, must be a comoving distance that takes into account 
time dilation along the path of GWs from the source at an earlier time in the universe,
\begin{equation}
D_c = c \int^{t_0}_{t_{\mathrm{src} }}{\frac{dt}{a(t)}},
\end{equation}
where $a(t)$ is the scale factor of the expanding universe at time $t$, $t_0$ denotes the observation time, and $t_{\mathrm{src}}$ denotes the emission time at the source. 
Because $f$ and L are present-day observed quantities in Eq.
(\ref{Eq:GeoCrit}), it is the \textit{comoving} distance that is encoded in the pulsar timing residuals measured at Earth.
Another way to see this is to recognize that the GW phase, $\Phi$, is the relevant quantity governing the timing residual, and the GW phase, $\Phi \propto \int{f (1+z) dt}$, where $f(1+z)$ is the GW frequency at redshift $z$ along the path. Since $a(t)=(1+z)^{-1}$, then $\Phi \propto \int{\frac{dt}{a(t)}} \propto D_c$.

%%%%%%%%%%%%%%%%%%%%%%%%%%%%%%%%%%%%%%%%%%%%%%%%
%%% FIGURE %%%
%%%%%%%%%%%%%%%%%%%%%%%%%%%%%%%%%%%%%%%%%%%%%%%%   

\subsubsection{Mathematical Argument}
\label{S:Mathematical Argument}
Following DF11, we outline the derivation of the pulse arrival time correction due to curved GW wavefronts. Building upon our previous setup, we illustrate relevant quantities for the derivation in panel b) of Figure \ref{Fig:schem}.

Assuming that the pulsar emits regular pulses at a much higher frequency than that of
the passing GW \citep[however, see][]{AngelilSaha:2015}, the extra light travel time of a given
pulse due to the passing GW, the arrival time correction, is given by
\citep{FinnLommen:2010, Anholm+2009, BookFlanagan:2011},
\begin{eqnarray}
t_{\GW}(t) &=& \frac{1}{2} \hat{n}^i \hat{n}^j \H_{ij} 
\label{Eq:taudef} \\
\H_{ij} &=& \frac{L}{c} \int^0_{-1} h_{ij}\left[ t+\frac{L}{c}\xi, 
-L\xi\mathbf{\hat{n}} \right] d\xi ,
\label{Eq:Hij1}
\end{eqnarray}
which is the integral of each component of the transverse-traceless GW metric
perturbation $h_{ij}(t, \mathbf{x})$ along a null geodesic connecting the pulsar and the Earth, parameterized by $\xi$
\footnote{
	Note that for our purposes it is sufficient to compute this pulse travel-time correction due to the intervening GW. Whereas the `timing residual' often encountered in the literature \citep[\eg,][]{McGrathCreighton:2020}, is this quantity divided by the pulsar period and integrated over the observation time. 
}. 
Here $\mathbf{P}$ is the position of the pulsar and $\mathbf{\hat{n}}$ is the unit vector pointing from Earth to the pulsar.

Writing out the strain as a function of the spacetime coordinates in the source frame, and in terms of the Fourier components,
\begin{eqnarray}
&&h_{ij}(t, \bX) = \\ 
&=& \frac{1}{|\bX - \bS|} \int^{\infty}_{-\infty}{\left[ \tA_{ij}(f_s,\hk) e^{2 \pi i f_s |\bX - \bS|/c}\right]  e^{-2\pi i f_s t} df_s d^3k}. \nonumber
\end{eqnarray}
where the $s$ subscript labels the source frame with no red-shifting of the frequency. Here $\bX$ is the vector pointing from Earth to the EM pulse wavefront that is traveling the null geodesic connecting the Earth and pulsar, $\bX = L\xi \hn$. $\bS$ is the vector pointing from the Earth to the source, $\bS = D_c \hs$, where $D_c$ is the comoving distance to the source. The quantities $\tA_{ij}$ are the components of the wave strength.

The spherical GW wavefront propagates along the unit vector,
\begin{equation}
\hk = \frac{\bX - \bS}{|\bX - \bS|}.
\end{equation}
The magnitude $|\bX-\bS|$, namely the distance from source to wavefront at point $\bX(\xi)$, can be found by the law of cosines. For a flat universe in comoving coordinates,
\begin{equation}
r(\xi) \equiv |\bX - \bS| = D_c \left[1 + \left(\frac{L\xi}{D_c}\right)^2 + 2 \frac{L \xi}{D_c}\cos{\theta} \right]^{1/2}.
\label{Eq:roxi}
\end{equation}
Note that the GW amplitude is proportional to $\tA_{ij}/r(\xi)$, which reduces to $\tA_{ij}/D_c$, as expected for $D_c \gg L$.

Combining Eqs. (\ref{Eq:taudef}-\ref{Eq:roxi}), keeping terms to $\mathcal{O}(L/D_c)$, integrating, and separating terms in order of $L/D_c$, we find the expression for the plane-wave and first-order-curvature pulse travel-time corrections in Fourier space, matching Eq. (10) of DF11. To write their expression in a more elucidating form, we define
\begin{eqnarray}
\Delta T &\equiv& 
\frac{\left[\hnc^i \hnc^j \tA_{ij}\right]_s}{2 \pi f_o D_c} e^{2 \pi i f_s D_c/c} \nonumber \\
&\propto& \frac{\Mc^{5/3}_s f^{2/3}_s}{2 \pi f_o D_c} e^{2 \pi i f_s D_c/c} \mathcal{Q}(\alpha_P, \beta_P, \phi, \phi_0, I, \psi) \\ \nonumber
\label{Eq:tauGW1}
\end{eqnarray}
where in the last line we write out the $\tA_{ij}$ dependence assuming circular binary orbits. The function $\mathcal{Q}$ depends on the pulsar position angles $(\alpha_P, \beta_P)$, the binary orbital phase $\phi$ and phase reference $\phi_0$, the binary inclination $I$, and the polarization angle $\psi$. The other two angles of importance, the angular position of the GW source ($\alpha, \beta$), appear outside of $\Delta T$ through the angle $\theta$,
\begin{equation}
\cos{\theta} = \cos{\beta}\cos{\beta_P} \cos{(\alpha - \alpha_P)} + \sin{\beta}\sin{\beta_P}.
\end{equation}

Importantly, the denominator of Eq. (\ref{Eq:tauGW1}) includes the frequency at the Earth-pulsar detector. This is because the observed timing residual is set by the observed strain over the observed GW frequency. To put the above into observed quantities for sources at cosmological distance, use that $\Mc_o = (1+z) \Mc_s$ and $f_o = (1+z)^{-1} f_s$. Therefore,
\begin{eqnarray}
\frac{\Delta T_o }{\mathcal{Q}(\alpha_P, \beta_P, \phi, \phi_0, I, \psi)} 
\propto \frac{ \Mc^{5/3}_o f^{-1/3}_o }{2 \pi  D_L } e^{2 \pi i f_0 D_L/c},
\end{eqnarray}
where the luminosity distance $D_L = (1+z) D_c$.

Using our definition of $\Delta T_o$, the Fourier travel-time correction, broken into plane-wave and first-order curvature parts, becomes
\begin{widetext}
\begin{eqnarray}
\tau_{\GW} &=& \tau_{\mathrm{pw}} + \tau_{\mathrm{cr}} = 
 \frac{\Delta T_o}{2} \left \{ \exp\left( 2 \pi i \frac{f_o L}{c} \sin^2\frac{\theta}{2}\right) \frac{\sin\left( 2 \pi \frac{f_o L}{c} \sin^2\frac{\theta}{2}\right)}{\sin^2\frac{\theta}{2}} + \right. \nonumber \\
 &+&  \left. 2 (1+\cos\theta) \exp\left[ \pi i \frac{f_o L}{c} \left(4 \sin^2\frac{\theta}{2}  +   \frac{L}{2D_c}\sin^2\theta\right)\right]  
 \frac{\sin\left( \frac{\pi f_o L^2}{2 c D_c} \sin^2\theta\right)}{ \sin^2\theta }  \right\} .
 \label{Eq:tauGW}
\end{eqnarray}
\clearpage
\end{widetext}
Both terms still have a dependence on the distance through the usual $1/D_L$ in $\Delta T_o$. However, the curvature term now has a dependence on $f_o L^2/D_c$. \textit{Because we can independently measure $f_o$ and $L$, this term introduces a way to measure the comoving distance separately from the luminosity distance.}

The form of the travel-time correction also confirms our simple geometric argument of the
previous subsection, that the curvature term decreases in importance as 
$\pi f_o L^2/(2cD_c) \sin^2\theta \rightarrow 0$. Indeed when $D_c \gg \frac{f_o L^2}{c}$, the wavefront curvature can be neglected. 
For values typical of near-future PTAs, the timing residual due to the wavefront curvature corrections will be of order the plane-wave residual when
\begin{equation}
\frac{\pi f_o L^2}{2 c D_c} = 0.5 \left(\frac{f_o}{\mathrm{yr}^{-1}}\right) \left(\frac{L}{\mathrm{10 kpc}} \right)^2 \left( \frac{D_c}{1 \mathrm{Gpc}} \right)^{-1}. 
\label{Eq:GeoCrit2}
\end{equation}
\textit{Note that the dependence on pulsar distance is quadratic.}  Hence, if
future PTAs can precisely time pulsars out to 10-20~kpc, then the comoving distance can be
probed through GW timing parallax, at the same sensitivity as required for detection of the plane-wave terms, out to Gpcs, for all conceivable
supermassive black hole binary mergers detectable by PTAs. If the GW system is high signal-to-noise, then even greater distances can be probed.
However, in addition to the criteria considered thus far, measurement of the wavefront curvature also requires a precise measurement of the pulsar distance, which we discuss below.

\section{Hubble Constant Measurement}
When both the luminosity distance and the comoving distance can be measured, and distinguished from each other, for the same binary GW source, the redshift and hence Hubble constant $H_0$ can be measured. To distinguish the two distances we require that the fractional errors on $D_L$ and $D_c$ be less than $(D_L - D_c)/D_c = z$,
which is $\sim0.25$ at 1~Gpc.
When this is possible, the redshift of the source can be recovered with uncertainty,
\begin{equation}
\delta z = \sqrt{ \left(\frac{\delta D_L}{D_c}\right)^2 + \left(\frac{D_L}{D_c}\frac{\delta D_c}{D_c}\right)^2 }
\label{Eq:dz}
\end{equation}
and can be used to measure the Hubble constant via,
\begin{equation}
H_0 = \frac{c}{D_c(z)} \int^{z}_0{\frac{dz'}{E(z')}},
\end{equation}
with relative uncertainty,
\begin{equation}
\frac{\delta H_0}{H_0} = \sqrt{ \left( \frac{\delta z}{ E(z) } \right)^2 
\left( \int^{z}_0 \frac{dz'}{E(z')}\right)^{-2}  +  \left( \frac{\delta D_c}{D_c} \right)^2  }.
\label{Eq:dH}
\end{equation}

\subsection{Distance Measurement Precision and PTA Dependence}
\label{S:Distance Measurement Precision and PTA Dependence}
We estimate the fractional error in the Hubble constant measurement
by considering PTA detections above a cutoff SNR and hence a constant fractional
error on $D_L$. As a fiducial value we use $\delta D_L/L \sim 10\%$ estimated
in \citep{CorbinCornish:2010} for detections with a signal-to-noise-ratio (SNR) of 20 (see \S \ref{S:Luminosity Distance Measurement}). Next, we numerically estimate the
precision in the $D_c(z)$ measurement.

\subsection{Main Challenges}
A requirement for the measurement of $D_c$ is that the pulsar coordinates are precisely known, to within approximately a GW wavelength. While this does not impose a stringent constraint on the measured precision of the pulsar angular coordinates, it does strongly constrain the required precision on the pulsar distance measurement (\eg, DF11, \citep{CorbinCornish:2010}).

This $\delta L$ requirement is seen geometrically from panel a) of Figure \ref{Fig:schem}. 
Imagine that the pulsar is at distance $L$ and angle $\theta=\pi/2$. In the plane-wave limit the GW phase observed at Earth is $\Phi = \Phi_0 + 2 \pi f \left( t - L/c\right)$, while for spherical wavefronts the phase is approximately $\Phi = \Phi_0 + 2 \pi f \left[ t - L/c\left(1 - L/D_c\right)\right]$. Hence, if $L$ is not known to within $\delta L \lesssim (L/D_c)L$, which is approximately $\lambda_{\GW}$ in the limit of Eq. (\ref{Eq:GeoCrit}), one cannot distinguish between a phase difference due to a different $L$ or due to the $L/D_c$ correction. In Figure \ref{Fig:schem} this translates to a degeneracy between moving the pulsar to a different distance in the plane-wave approximation, or keeping it fixed but with differently curved wavefronts (different $D_c$).

One might expect a similar requirement exists for $D_L$, as it also relies on including combined phase information at the pulsar and at Earth. However, this requirement can be circumvented by simultaneously fitting for the pulsar distances in a joint analysis with the binary properties and $D_L$. The timing residual from each pulsar is modulated at the beat frequency between the GWs at the Earth and at the pulsar, which encodes the chirp from which $D_L$ is measured. As long as a residual is monitored for long enough to measure a beat modulation (the criteria of Eq. \ref{Eq:DLcond}), the chirp can be measured from $\dot{f} \propto c f_{\mathrm{beat},i}/L_i$, for the $i^{\mathrm{th}}$ pulsar. Hence simultaneous fitting for the $L_i$ will yield the one value of $\dot{f}$, and so $D_L$, to higher precision with more pulsars. See also the discussion in \cite{CorbinCornish:2010, Ellis:2013}.

We first demonstrate how the $H_0$ measurement depends on pulsar distance
uncertainties and the number of pulsars in the array, we then turn to a discussion of the practicality of such measurements and possible avenues towards making them a reality.

\subsection{PTA Dependence}
We envision an idealized PTA with $N_p$ pulsars having randomly drawn angular
coordinates ($\alpha_i, \beta_i$) on the sky, at randomly drawn distances
$L_i$ in the range $L_{\min}$ to $L_{\max}$, and with
fractional distance errors measured in units of the gravitational wavelength $\lambda_{\GW}$, $\delta L/L = \chi \lambda_{\GW}/L_{\max}$. We consider further that, for each pulsar, the
timing residual divided by the prefactor $\Delta T_o$ in Eq. (\ref{Eq:tauGW})
can be measured to within a constant fractional error of $\delta \tau /\tau$.
This essentially subsumes errors on the remaining binary parameters into
$\delta \tau_i$ and will generally be dependent on the SNR.

We generate mock observed timing residuals by calculating the expected travel
time correction, $\tau_i$, from Eq. (\ref{Eq:tauGW}). We draw observed values
$\tau_{\obs,i}$ from a normal distribution with mean and standard deviation
given by $\tau_i$ and $\delta {\tau_i}$, respectively. We also draw observed
pulsar distances $L_{\obs,i}$ from a normal distribution with mean and
standard deviation given by $L_i$ and $\delta {L_i}$, respectively. 
We recover the source parameters from the observed arrival time deviations by
minimizing a least squares statistic that compares to the model, Eq.
(\ref{Eq:tauGW}), but with $L_{\obs,i}$ as the input pulsar distances,
\begin{equation}
\sum^{i=N_p}_{i=1}\left(|\tau_{\GW}(D_c, \alpha, \beta, L_{\obs,i})| - |\tau_{\obs,i}| \right)^2,
\end{equation}
where $|\cdot|$ denotes the norm of the complex Fourier timing deviations. We 
impose a log-uniform prior on $\log{D_c/\mathrm{Mpc}} \in \left[0,
5\right]$ and uniform priors on the angular source coordinates $\alpha \in
\left[0, 2\pi\right], \beta \in \left[0, \pi\right]$.  For fixed
$\tau_{\obs,i}$, we carry out 100 such minimizations for 100 different
realizations of the $L_{\obs,i}$. We quote the mean and standard deviation of
the 100 sets of resulting source parameters as estimates for the recovered
parameters and their uncertainties.

Throughout, we consider a fiducial PTA with $N_p$ pulsars in the
$L_{\min}=1$~kpc to $L_{\max}=20$~kpc distance range and a error on $L$
parameterized in units of GW wavelengths. The fractional error on $\tau$ will
be SNR dependent; for the purpose of this study, we choose a fiducial value of
$10\%$. We study the affect of varying these choices below.

For computational purposes, we draw pulsar sky locations within $\pi/4$ from
the optimal $\theta=\pi/2$. This means that an isotropic pulsar distribution
would require twice the number quoted here, though, for a favorably positioned
source, a pulsar distribution biased by the Milky Way plane would require less
pulsars than our $N_p$ suggests. We find below that a factor of a few in our
predicted pulsar numbers is not significant compared to other uncertainties
and may not be a limiting issue given that 100's to 1000's of pulsars may make
up future PTAs \citep{Smits+2011}.  Finally, we consider a fiducial GW source
frequency of $f_o = 10^{-7}$~Hz, where the GW parallax distance
determination is most effective.

%%%%%%%%%%%%%%%%%%%%%%%%%%%%%%%%%%%%%%%%%%%%%%%%
%%% FIGURE %%%
%%%%%%%%%%%%%%%%%%%%%%%%%%%%%%%%%%%%%%%%%%%%%%%%    
\begin{figure*}
\begin{center}$
\begin{array}{c c c}
\includegraphics[scale=0.35]{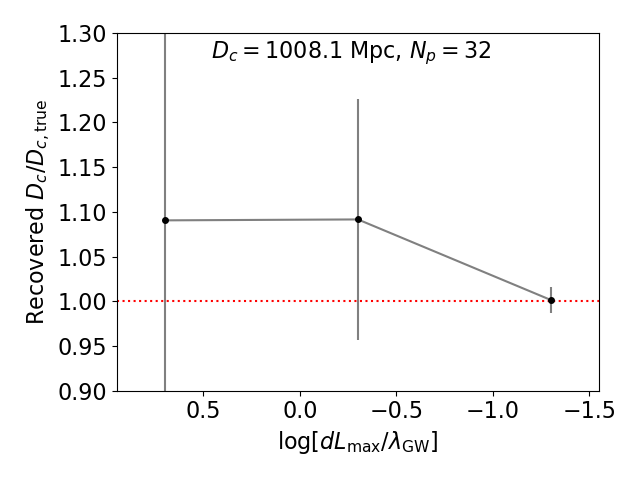} &
% %
\includegraphics[scale=0.35]{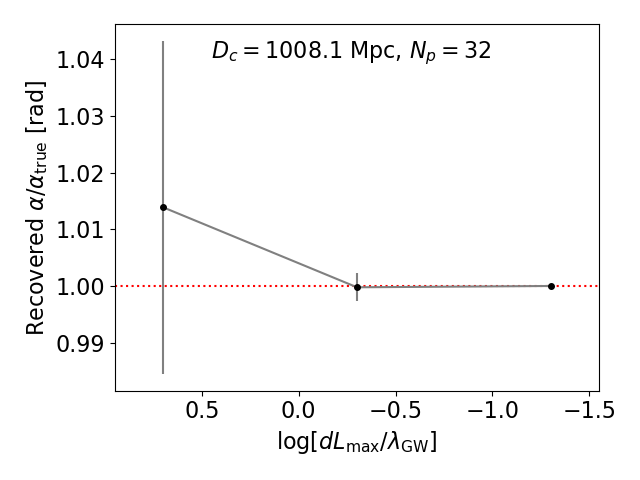} &
% %
\includegraphics[scale=0.35]{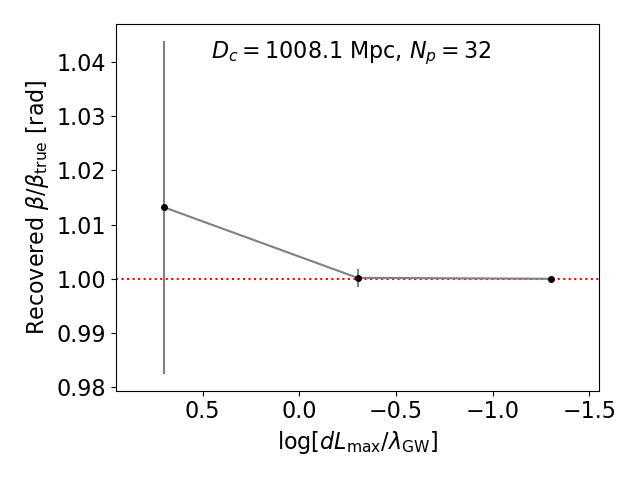} \\
\includegraphics[scale=0.35]{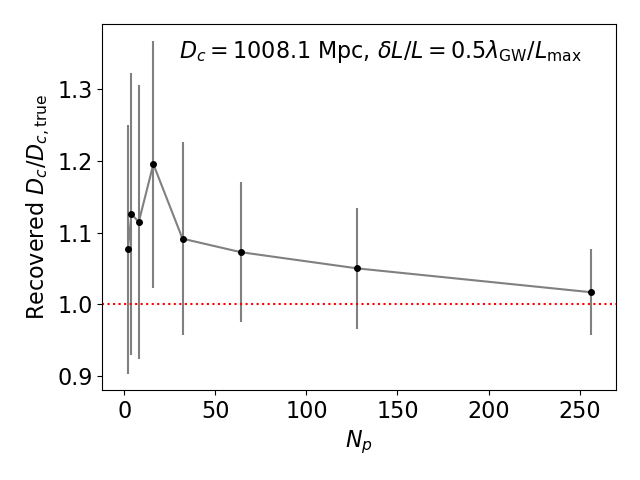} &
\includegraphics[scale=0.35]{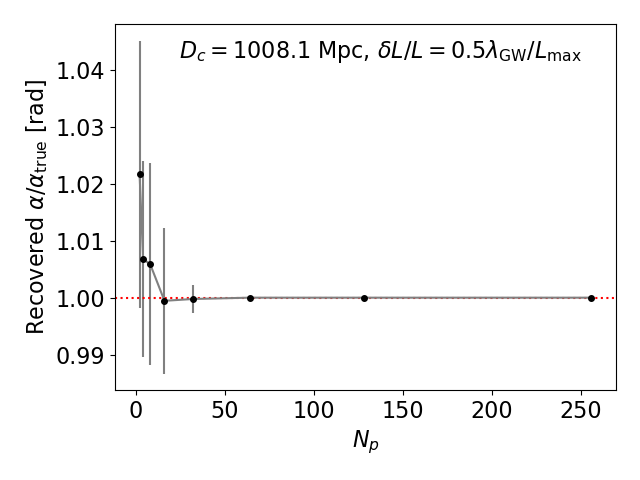} &
\includegraphics[scale=0.35]{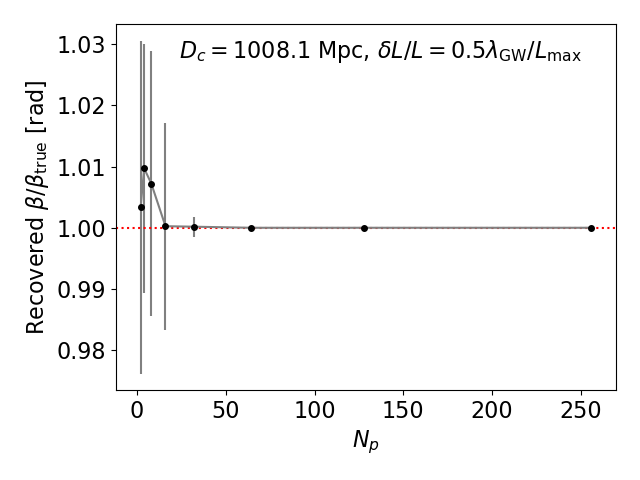} 
\end{array}$
\end{center}
\vspace{-10pt}
\caption{
The recovered comoving distance, $D_c$, and sky coordinates, ($\alpha, \beta$), of the GW source for PTAs with varying numbers of pulsars and precision in the distance measurement
of these pulsars. The considered PTA assumes pulsars lying between 1 and 20~kpc from Earth and with $\delta \tau/\tau=0.1$, and $dL_{\max}$ refers to the worst precision in the array, on the most distant pulsars.
}
\label{Fig:npdL}
\end{figure*}

Figure \ref{Fig:npdL} shows how well our fiducial PTA recovers the comoving
distance (left) longitude (middle), and latitude (right) of our GW source when
it is placed at a redshift of $z=0.25$ with angular positions
($\alpha,\beta$)=($\pi/4,\pi/4$). The top row shows that indeed the discussed
requirement on the pulsar distance is borne out in our numerical propagation
of errors experiment. Only for $\delta L \lesssim 0.5 \lambda_{\GW}$ is a
constraint made on the source parameters. For a 32 pulsar array, $D_c$ is
constrained at the $10\%$ level for $\delta L/L = 0.5 \lambda_{\GW}/L_{\max}$,
while the angular coordinates of the source are constrained to the sub-$1\%$
level. These source constraints tighten proportionally to $\delta L$.
In the bottom row, we vary the number of pulsars in the $\delta L/L = 0.5 \lambda_{\GW}/L_{\max}$ array. We find that the source coordinates are poorly
constrained for $N_p \leq 16$, but accuracy and precision of parameter recovery increase with increasing pulsar number. Increasing the number of pulsars for a PTA with $\delta L \geq \lambda_{\GW}$ does not allow a better (or any) measurement of the source parameters.

\subsection{Precision of Redshift and Hubble Constant Measurement}
\begin{figure*}
\begin{center}$
\begin{array}{c c}
\includegraphics[scale=0.53]{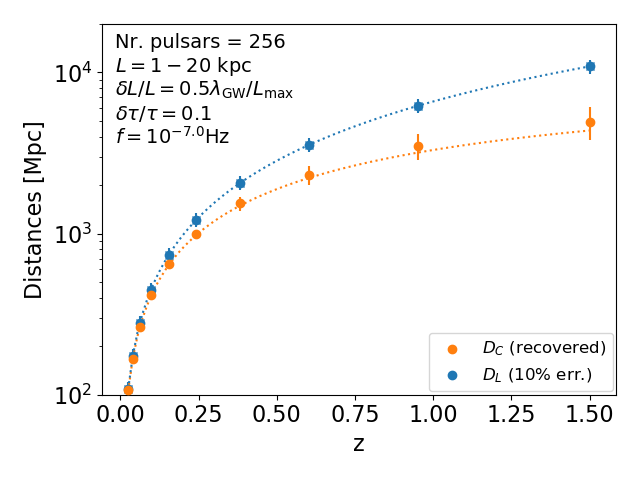}  &
\includegraphics[scale=0.53]{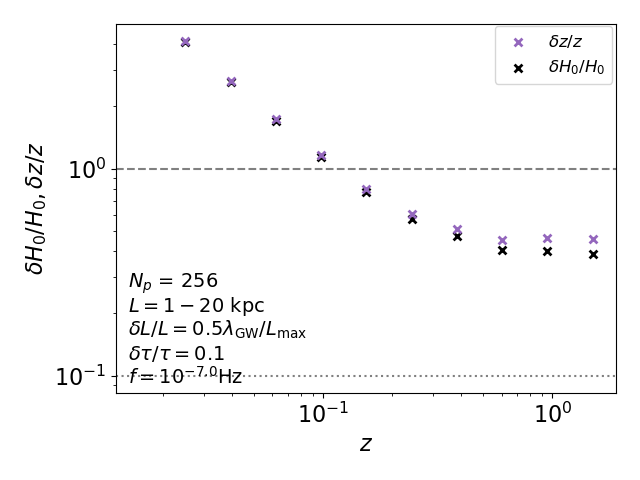} 
\end{array}$
\end{center}
\vspace{-10pt}
\caption{
{\em Left:} Recovery of comoving and luminosity distances, $D_c$ and $D_L$, vs. redshift, $z$, for a PTA with 256 pulsars, with $\delta L/L=0.5 \lambda_{\GW}/L_{\max}$. {\em Right:} the corresponding fractional errors in the measured redshift and the Hubble constant. The fractional error in $D_L$ is assumed constant.
}
\label{Fig:zHerr}
\end{figure*}

 In the left panel of Figure \ref{Fig:zHerr}, we plot recovered comoving
distances as a function of redshift (orange) for fiducial PTA and source
properties and using 256 pulsars. For reference, we also plot the
corresponding luminosity distances with $10\%$ fractional errors (blue). The
dotted lines show the theoretical expectation for each distance
measure\footnote{We use $\Omega_M=0.3$, $\Omega_{\Lambda}=0.7$, and $h=0.7$
throughout.}. Accurate determination of luminosity and comoving distances,
and hence determination of the source redshift, is possible when recovered
values are consistent with the theoretical values, and when the blue and
orange error-bars do not overlap.

The right panel of Figure \ref{Fig:zHerr} uses the distance errors in the left
panel and Eqs. (\ref{Eq:dz})-(\ref{Eq:dH}) to display the fractional error
in the redshift and the Hubble constant measurements for the corresponding
points in the left panel. For this specific GW source ($\alpha=\beta=\pi/4$,
$f=10^{-7}$~Hz) and PTA, we find that the redshift and Hubble constant can be
determined to better than order unity for $z \gtrsim 0.1$ and to within $40\%$
for $z \gtrsim 0.5$. At $z\leq0.1$, $D_c$ and $D_L$ are indistinguishable from
each-other, but their measurement could still impose upper limits on $z$ and
$H_0$. Choice of smaller $\delta D_L/D_L=0.01$ decreases the redshift at
which a determination of $H_0$ could be made, bringing the low-$z$ side of the
curve in the right panel of Figure \ref{Fig:zHerr} down to the level of the
high-$z$ values, allowing $30-40\%$ fractional errors on $H_0$ and
$z$ for $z\lesssim1.5$.

\begin{figure*}
\begin{center}$
\begin{array}{c c}
\includegraphics[scale=0.53]{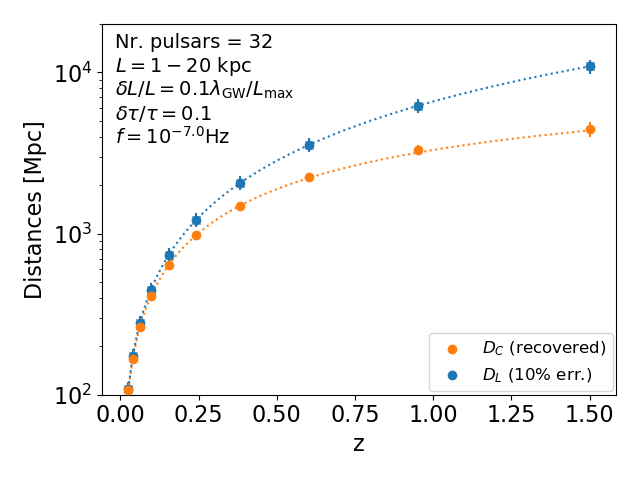}  &
\includegraphics[scale=0.53]{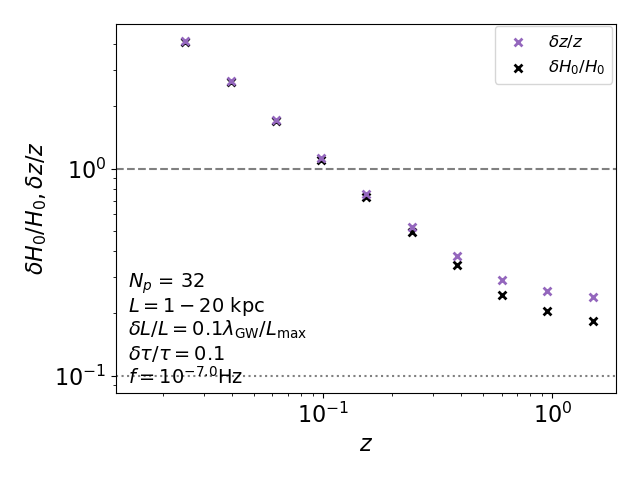} 
\end{array}$
\end{center}
\vspace{-10pt}
\caption{
Same as Figure \ref{Fig:zHerr}, but for $5\times$ more precise pulsar distances but $8 \times$ fewer pulsars. Precise pulsar distances are more important than the number of pulsars in the array.
}
\label{Fig:zHerrBest}
\end{figure*}

To further demonstrate the dependence on PTA properties, Figure
\ref{Fig:zHerrBest} replicates Figure \ref{Fig:zHerr} but now for a more
optimistic scenario where the pulsar distances can be measured to a 5 times
higher precision of $\delta L/L = 0.1 \lambda_{\GW}/L_{\max}$, but including
eight times fewer pulsars ($N_p=32$). In this case, the minimum redshift
required for order-unity-precision distance measurements remains at $z=0.1$,
but $\leq20\%$ level measurements of the Hubble constant are possible for
$z\geq1$. We do not consider higher redshifts as such binary GW sources are
likely not detectable beyond this range with Square-Kilometer-Array (SKA)-era
PTAs \citep{SVV:2009}, though futuristic arrays may extend beyond these
redshifts. Fractional errors on $H_0$ and $z$ would decrease further for
higher frequency sources, a better measurement of $D_L$, or the inclusion of
more well-measured pulsars.

%%%%%%%%%%%%%%%%%%%%%%%%%%%%%%%%%%%%%%%%%%%%%%%%
%%% FIGURE %%%
%%%%%%%%%%%%%%%%%%%%%%%%%%%%%%%%%%%%%%%%%%%%%%%%    

%%%%%%%%%%%%%%%%%%%%%%%%%%%%%%%%%%%%%%%%%%%%%%%%
%%% FIGURE %%%
%%%%%%%%%%%%%%%%%%%%%%%%%%%%%%%%%%%%%%%%%%%%%%%%    

\subsection{Beyond the Pulsar Distance Constraint}

Up until now we have only shown the dependence of the $H_0$ measurement on the
parameters of a hypothetical PTA. Such a hypothetical PTA, however, requires
pulsar distance uncertainties, $\delta L \lesssim \lambda_{\GW}$, that would
pose a great challenge to achieve with present methods. This is because the GW
wavelength of interest is $\lambda_{\GW} = 0.1-1.0$~pc for a
$10^{-7.0}-10^{-8.0}~\mathrm{Hz})^{-1}$ frequency range, which amounts to a
precision on the pulsar distance that has been approached only for a few
nearby pulsars \citep[\eg,][]{Deller_1pcDisterr+2013, Mingarelli+2018}. In the
near future, precise pulsar distance measurements using parallax from VLBI
determined astrometry will be able to achieve parallax angle uncertainties of
$\delta p = \mu as$ resulting in $\delta L \sim \delta p / p^2$ distance
measurements \citep{RiojaDodson_rev2020}. For $L=1$~kpc, this is $\delta L
\sim 1 \mathrm{pc} (L/(1~\mathrm{kpc}))^2$. Hence, with astrometry limited to
the $\mu as$ level and without another way around the pulsar distance problem,
the technique presented here would seem to be limited to using pulsars within
$1$~kpc, and so limits a comoving distance measurement to sources with
$D_c\lesssim100$~Mpc via Eq. (\ref{Eq:GeoCrit2}).

One way beyond this is to look to space VLBI, where sub-$\mu as$ astrometric
limits are conceivable \citep{EHTspace:2019, Gurvits_sVLBI:2018}. Though parallax uncertainties of $\sim10^{-2}~\mu as$ (sub-mm and higher frequency VLBI over $10^6$ km baselines) may be needed to achieve the required precision for pulsars out to $10-20$~kpc.

In addition to relying on future high-precision pulsar distance determinations, it may be possible to use GW detections with near-future PTAs to measure the pulsar distances and build a ladder of precise distance measurements from the well measured nearby pulsars out to the distant pulsars, and then to the GW sources by means of GW parallax. We speculate on a few possible scenarios that could lead towards this goal:
% \begin{enumerate}
	(\textit{i}) A calibrating GW source with known redshift could be used to measure the pulsar distances using the chirp and GW curvature for an entire array. The redshift could derive from localization of a host galaxy and/or an assumed cosmology coupled with a measured luminosity distance.
	(\textit{ii}) A nearby source of GWs ($D_c \leq 100$~Mpc) could allow measurement of the comoving distance and source location using well-measured nearby pulsars ($L\leq 1$~kpc). The known source location could be used to update the coordinates of distant pulsars, which are more greatly affected by wavefront curvature terms.
% \end{enumerate}
%
A rigorous analysis of these possibilities and, in general, the joint recovery of the luminosity, comoving, and pulsar distances is the subject of current and future work.

\section{Discussion and Conclusions}

We have demonstrated that PTAs can measure both the luminosity distance and
the comoving distance to a subset of resolved binary sources of GWs. In doing
so, they can measure the source redshift and the Hubble constant. Thus the
PTAs, by themselves, could become cosmological instruments. The distance out
to which such a measurement can be made, however, depends on the level to
which pulsar distances can be determined.

Currently, PTAs are operating with tens of pulsars at distances out to a few
kpc. Distance errors range from a fraction of a percent to order unity, with
some reaching down to the sub-pc level precision required to measure a
comoving distance as described here \citep[\eg,][]{Deller_1pcDisterr+2013,
Mingarelli+2018, Deller57PulsarDsts+2019}. In the coming years, the
SKA is expected to expand the pulsar population drastically
\citep{Janssen+2015}. Ref. \cite{Smits+2011} estimates that $\sim9000$ pulsars
are detectable by the SKA out to $\mathcal{O}(10)$~kpc with better than $20\%$
error on their distances. If a few of these are suitable for high
precision timing and also high precision distance measurements, either with
future space-VLBI, or via GW-based pulsar distance measurements, then the PTAs
envisioned here, with 10's to 100's of pulsars between 1 and 20~kpc and with
$\lesssim$pc distance errors, could be realized, though most likely in the post-SKA
era. Note that Ref. \citep{Smits+2011} considers only galactic pulsars (see
their Fig. 1); pulsars in the Magellanic clouds could provide MSPs out to
$40$-$60$~kpcs \citep{CrawfordKasp:MagellPulsars2001, Manchester:Magell+2006,
Ridley:LMCpulsars:2013, TitusToonen:SMCpulsars+2020}. Though, again,
without ultra-high precision distance measurements, such distant pulsars may
only be useful for measuring the luminosity distance, and not the comoving
distance. In addition to the SKA, the next generation Very Large Array
\citep[ngVLA,][]{ngVLA:GWs:2018} and astrometric pulsar distance measurements
with WFIRST \citep{WFIRST_astrometry:2019} will make the pulsar-distance
errors envisioned here even more feasible, at least for nearby, $L
\lesssim 1$~kpc, pulsars. Such future arrays will also likely decrease the
expected error in the luminosity distance measurement, and further reduce the
error in the described Hubble constant measurement.

If the Hubble constant can be measured in this manner for tens of sources,
then a PTA-only measured value could reach a few to $10\%$ precision. While
the number of such resolved `foreground' binaries that will be detected is
uncertain and relies on the poorly constrained binary population, multiple
studies have attempted to estimate this number. Ref. \cite{Kelley_SS+2018}
estimates that it may indeed be the resolved single binary sources that are
detected before a stochastic GW background, with a detection every few years.
Older models suggest that the number of such detections may be an order of
magnitude lower \citep{SVV:2009,Ravi:2014, Rosado:2015}. Reassuringly, Ref.
\cite{SVV:2009} shows that the most probable redshift range for resolved
sources is between $0.2\lesssim z\lesssim 1.5$, in the right range for the
measurement envisioned here.  However, \cite{Kelley_SS+2018} shows that while
the resolved single sources are the most common at the higher GW frequencies
considered here $\sim10^{-7}$~Hz, their amplitudes are lower and the PTAs are
less sensitive at these frequencies. This leads Ref. \cite{Kelley_SS+2018} to
conclude that the optimal single-source {\em detection} frequencies for near
future PTAs lie at $\sim10^{-8}$~Hz. At these lower frequencies, the GW
parallax measurement is more difficult, as the curvature term is less
important for a GW binary at the same distance (Eq. \ref{Eq:GeoCrit2}),
however, at lower frequencies the requirement on the pulsar distance
uncertainty is lessened by the same amount. Future work could analyze expected
supermassive black hole binary populations in light of the measurement at
hand, quantifying how many sources per redshift and frequency will contribute
to a meaningful measurement of the Hubble constant.

For nearby GW sources ($z\lesssim0.1$), we found that only upper bounds on the
redshift and Hubble constant can be set because $D_c$ and $D_L$ are within
$10\%$ of each-other. This is partly due to our adopted $10\%$ fractional
error on the luminosity distance measurement. If this can be improved upon,
as it very well could be by the time that the $D_c$ measurement is
feasible, then one can take advantage of more nearby sources. In addition,
because the comoving distance can be measured to high precision for these
nearby sources, its measurement could facilitate identification of the binary
host galaxy (as discussed in DF11), and allow a standard-siren-type
determination of the Hubble constant, as well as offer important astrophysical
insight into, \eg, the morphology of supermassive black hole binary host
galaxies.

While we have provided a proof-of-principle error estimation focusing on the
largest error sources, future work should consider more realistic parameter
estimation techniques, and the precision and accuracy to which all of the
binary parameters can be recovered jointly
\citep[\eg,][]{TaylorEllisGair:2014,ZhuWenHobbs+2015}. For example, by not
modeling the orbital geometry of the GW source, we do not include binary
inclination or GW polarization factors that would affect the degree with which
inclination and luminosity distance can be disentangled. Furthermore,
we have not included the frequency evolution of the binary when including the
wavefront curvature terms in the arrival time corrections, but this will be
necessary for joint recovery of the luminosity and comoving distances.
Finally, techniques that independently fit for the pulsar distances as part of
the model \citep[\eg,][]{Lee+ssPTA:2011, Ellis:2013} could enhance the
precision of source parameter recovery presented here and should also be
considered for application to cosmology with PTAs
\footnote{During the review
process of this work, \citep{McGrathCreighton:2020} posted a preprint
detailing recovery of binary GW source parameters including a distance using
the curvature of GW wavefronts while including the changing GW frequency across
the Earth-pulsar baseline. No work has yet differentiated the comoving and
luminosity distances in such an analysis.}.

In summary, we have presented a novel method by which to measure the Hubble
constant without the use of EM radiation, by assuming only general relativity,
and without the need to model astrophysical properties of the emitting source
of GW radiation. This measurement can be made uniquely by future PTAs that can
determine pulsar distances in the array to sub-pc precision. Depending on the
distance to pulsars for which such a distance measurement can be made, this
would result in a single-source determination of the Hubble constant at the
tens of percent level at redshifts $0.1\lesssim z \lesssim1.5$. Tens of such
detections could yield a $\lesssim 10\%$ measurement of the Hubble constant
from gravitational signals from cosmological sources.

\acknowledgements
We thank Matthew C. Wilde, Stephen Taylor, Luke Kelley, Casey McGrath, Julian
Creighton, Zoltan Haiman, and attendees of the October 23, 2020 NANOGrav
Astro-WG meeting for useful discussions during the preparation of this work.
Financial support was provided through funding from the Institute for Theory
and Computation Fellowship (DJD) and through the Black Hole Initiative which
is funded by grants from the John Templeton Foundation and the Gordon and
Betty Moore Foundation.

\bibliography{refs}
\end{document}